# Molecular Dynamics Simulations of Cytochrome c unfolding in AOT Reverse Micelles: the first steps

*Stéphane Abel[+,\*], Marcel Waks[‡,§] and Massimo Marchi[+]*

[+]Commissariat à l'Energie Atomique, DSV/iBiTecS/SB2SM, CNRS URA2096, Gif sur Yvette, F-91191, France; [‡]UPMC, Laboratoire Imagerie Paramétrique, F75006 Paris, France [§]CNRS, LIP UMR7623, F-75006, Paris, France.

[\*]To whom correspondence should be addressed. E-Mail: stephane.abel@cea.fr

This paper explores the reduced form of horse cytochrome c confined in reverse micelles (RM) of sodium bis-(2-ethylhexyl) sulfosuccinate (AOT) in isooctane by molecular dynamics simulation. RMs of two sizes were constructed at a water content of $W_o = [H_2O]/[AOT] = 5.5$ and 9.1. Our results show that the protein secondary structure and the heme conformation both depend on micellar hydration. At low hydration, the protein structure and the heme moiety remain stable, whereas at high water content the protein becomes unstable and starts to unfold. At $W_o = 9.1$, according to the X-ray structure, conformational changes are mainly localized on protein loops and around the heme moiety, where we observe a partial opening of the heme crevice. These findings suggest that within our time window (10 ns), the structural changes observed at the heme level are the first steps of the protein denaturation process, previously described experimentally in micellar solutions. In addition, a specific binding of AOT molecules to a few lysine residues of the protein was found only in the small-sized RM.

**I. Introduction**

Reverse micelles (RM) are stable, isotropic, water microemulsions suspended in organic solvents[1]. One of the most described RM systems is formed by sodium bis-(2-ethylhexyl) sulfosuccinate (AOT) and water in isooctane. This ternary system has been widely studied in last decades by various physical methods (see refs. [2-4] for data and references). The size of the water pool (i.e. the water content of the system) can be varied from 10 to 100 Å by changing $W_o$, the water-to-AOT molar ratio: $[H_2O]/[AOT]$ [5]. In many cases, the secondary structure of confined proteins is conserved in these systems which are considered as a biomimetic microenvironment and provide thus a simple model for the study of the restricted microenvironment effects on the structure and dynamics of confined peptides and proteins [6-8].

Recently, molecular dynamic simulations (MD) have been successfully used to gather additional information about AOT RM structure and dynamics at the atomic level [9-16]. These studies have underlined the effect of the charged, restricted microenvironment on the confined water structural [13,14,16] and dynamical properties [13,14,17] existing in RM. More recently molecular simulations have been extended to examine the influence of hydration on peptides and small proteins confined in AOT surfactant RM [17] or in fluorosurfactants [18]. In the former study, we have examined the structural changes of an alanine octapeptide ($A_8$) in its canonical α-helix conformation, in small-sized AOT RM at $W_o < 7.0$. We have shown that the secondary structure stability of the confined peptide depends on the water pool size, the surfactant headgroup and the peptide backbone polar groups hydration balance, as suggested by recent experimental [19] and theoretical studies [20].

Cytochrome c (CYTC) is a small (104 residues), globular, basic (pI = 10.5) hemoprotein associate with the inner membrane of the mitochondrion. It plays a crucial role in the electron transfer (ET) within the respiratory chain [21] and in other cellular mechanisms, such as apoptosis [22]. According to the STRIDE program [23], the protein displays 5 α-helices ($α_{1-5}$), 2 β-sheets ($β_{1-2}$) and 4 major loop ($L_{1-4}$) domains (Figure 1). Most of the positively charged residues are located around the heme crevice [24]. Depending on the microenvironment, CYTC can adopt a number of conformational states. For example, membrane-bound CYTC has been thoroughly investigated [25-27]. It was shown that the membrane anionic environ-ment disrupts the native protein structure, leading to molten globule-like states [26,28]. In such states the absence of the characteristic heme absorbance band at 695 nm, is paralleled with the disruption of the Met-80 ligation to the iron atom [29,30]. In AOT reverse micelles, it was suggested that the protein is located at the micellar interface, with its positively charged residues close to the AOT head groups [31,32]. As in membranes, electrostatic interactions also occur between the AOT anionic polar head groups and the positively charged protein residues leading to identical spectral characteristics, reported at different RM water contents [29,30]. Such results have been interpreted by the opening of the CYTC heme crevice and the successive disruption of the Met80-Fe and His18-Fe bonds under the electrostatic field. However, it was also pointed out [32,33] that despite the possible absence of the Met80-Fe ligation to the iron atom, the secondary structure of the protein remains close to the native state at $W_o \leq 10$, whereas it becomes unstable with increasing values of $W_o$, displaying a lower α-helix content. To examine at the atomic level these phenomena, we have extended our previous atomistic model of AOT RM with confined peptides, to explore the conformation changes of CYTC in AOT RM at $W_o = 5.5$ and 9.1. We have compared these results with those of the protein in bulk water or in the crystal.

This paper is organized as follows: in the next sections, we will describe the preparation procedure of the systems and the potential used to model these systems. It will be followed by a section covering the MD procedures, the results and interpretations of our simulations.

**II. Methods**

For this study, the protein starting structure consists of the X-ray structure of the horse cytochrome c (PDB: 1HRC) refined at 1.94 Å by Bushnell et al. [34] (Figure 1). The protein contains 104 residues with an acetylated N-terminus and a net charge of +6. Three MD were performed with the reduced form of CYTC. Two simulations (CW) were carried out for the protein in water solution. In the next two, designated as CW-M1 and CW-M2, the protein was solvated inside RM of different sizes, at $W_o = 5.5$ and 9.1, respectively. Constructions and simulations of these systems are described below.

**A. Cytochrome c in water.** The preparation of the solvated protein (CW) simulation was made as follows: firstly, the crystal structure of the protein was minimized in vacuum with a few hundred steps of conjugated gradient, at a spherical cut-off of 10 Å. Next, the protein was inserted in a truncated octahedral box (with a=b=c=65 Å and α=β=γ=109.472°) containing 6681 molecules of TIP3 [35] water molecules, with 6 chloride ions to neutralize the system. The force field used to model the protein was based on CHARMM [36]. We note that in the original CHARMM force field, the heme is modeled in its *reduced* state (FeII) and several parameters around the heme moiety are missing (such as Met(S)-Fe and His(N)-Fe bonds). In consequence it does not reproduce accurately geometries obtained from DFT calculations or experimental vibrational spectra [37]. To deal with these problems, Autenrieth et al. [38] have recently developed new force fields in CHARMM and AMBER framework, for both oxidized and reduced states of heme to simulate more realistically the CYTC X-ray structure and the water-heme interaction energies. In this force field, the His-18(N) and Met-80(S) atoms are explicitly bonded to Fe atom. Since experimental results suggest these bonds are absent in AOT RM [30,31], the Autenrieth parameters cannot be used in the present study.

The system was then slowly heated from 0 K to 300 K by rescaling the atomic velocities to attain T = 300 K after 150 ps and by keeping the protein and the heme (including Fe ion) atoms frozen. An additional equilibration in NVE ensemble at T = 300 K with all the atoms free was also performed during 500 ps. After this step, the system was equilibrated in NVT (T = 300 K) during 500 ps and simulated in the NPT (P = 0.1 MPa and T = 300 K) ensemble for 10 ns after an additional equilibration of 200 ps. To compare the influence of the force field and particularly the Met(S)-Fe and His(N)-Fe bonds on the solvated protein structure and stability, we have also carried out an additional MD (CW-b) in the same conditions as described above with CW simulated using the Autenrieth parameters.

**B. Reverse Micelle Construction and Equilibration.** For the RM system we have adopted an all atom approach to model the AOT, water and isooctane molecules. Briefly, the force field and topology for the surfactant were derived from the CHARMM27 all atom parameters for lipids [39] and adapted for an AOT/water system [40], previously used to simulate AOT RM in isooctane [9]. We have applied the TIP3P model [35] to describe water. For the isooctane molecule, the bonded and non-bonded parameters were also taken from the CHARMM27 force field and tested to reproduce crucial properties of liquid isooctane (such as density, translational diffusion and isothermal compressibility) at P = 0.1 MPa and T = 300 K (data not shown).

To construct the starting configuration of the two RM, we have first estimated the number of AOT molecules per spherical micelle ($n_{AOT}$) with a linear fit of the aqueous core radius of gyration of (i.e. AOT headgroups and water), $R_g^w$, obtained from SAXS ([AOT] = 0.1 M and [CYTC] = 0.01 M at 284 K) [29] (Figure 2). To reduce the system size, we choose two values of the water-to-surfactant molar ratio $W_o$ = [H$_2$O]/[AOT] of 7.5 and 10. For these sizes, the curve fit gives $R_g^w \approx$ 21.4 Å and 25.1 Å, respectively. Using a surface area of 54 Å$^2$ [29] per AOT headgroup ($S_{AOT}$) for $W_o$ in the 5 – 45 range, using the expression $n_{AOT}=(4\pi R_g^{w\,2})/S_{AOT}$, we obtain $n_{AOT} \approx$ 106 and 147 for $W_o$ = 7.5 and $W_o$ =10, respectively. In the next step, we constructed two spherical clusters of the desired radius $R_g^w$ (i.e. 21.4 and 25.1 Å) containing 795 and 1470 water molecules from a water final configuration, obtained from a 100 ps MD runs of 2000 TIP3 bulk water molecules in a cubic box simulated at T = 300 K and P = 0.1 MPa. The corresponding numbers of surfactant molecules (including Na$^+$ ions) in an extended conformation were placed by hand around the water sphere surface, tail pointing toward the exterior using the visualization program DS ViewerPro by Accelrys. We are aware that building preassembled reverse micelle does not constitute an ideal approach to simulate a RM system, even if it is a common practice in MD studies of micelles [36,41-47] or reverse micellar structures [9,16,48,49,50,51], compared to a random approach where the surfactant monomers are randomly placed in the simulation box [52-56]. But, as recently shown by [15], the micellization process in AOT RM is long (close to 60 ns or more) and computationally expensive when the aggregates are small and/or when the micellar environment is simulated in great detail to reproduce experimental data.

After this construction phase, the two RMs were equilibrated at 500 K with a spherical cutoff of 10 Å for 500 ps, during which the water and the AOT headgroups were blocked to randomize the hydropho-bic chains of the surfactant. Then, the minimized structure of the protein was randomly inserted into the center of the micellar water core and all water molecules within 2 Å of protein atoms were removed within the two RMs of $W_o$ = 5.5 and 9.1, obtaining the systems CW-M1 and CW-M2 respectively. The RM-protein complex was again minimized with a few thousands steps of conjugated gradient and a cut-off 10 Å. Subsequently, the two micelles were inserted into a truncated octahedral box (with a=b=c=105 Å and a=b=c=110 Å and α=β=γ=109.472° for the smallest and the largest micelle, respectively) filled with 2587 and 3345 molecules of isooctane and six chloride ions to ensure electroneutrality of the system. The number of isooctane molecules was chosen to simulate a well defined $L_2$ phase with a mass fraction of isooctane larger than 20 % [2,29,57]. Finally, the simulated systems, referred to as CW-M1 and CW-M2 through this paper, contained 106 and 147 molecules of AOT, 580 and 1339 molecules of water, corresponding to $W_o$ = 5.47 and 9.1, respectively (see Table 1 for the system composition). To equilibrate the solvated RMs, we have used three steps as described in Ref. [17]. In a first time, the isooctane solvent was relaxed for 400 ps at 400 K with the RM-protein complex frozen. Then, the entire system was frozen at 0 K and then monotonically heated up to attain T = 300 K after 300 ps. These two systems were then equilibrated in the NVE ensemble during 300 ps and in NVT (T=300 K) during 200 ps. After these steps, the two micellar systems were simulated in the NPT ensemble for 10 ns each after an additional equilibration of 100 ps (CW-M1) and 200 ps (CW-M2), respectively.

**C. Molecular Dynamics.** All simulations described in this paper were performed in an isobaric-isothermal (NPT) ensemble at T = 300K and P = 0.1 MPa with the sequential and parallel versions of the program ORAC [58]. To simulate in the NPT ensemble, ORAC uses a method based on the extended system approach [59-61]. This technique involves adding extra (virtual) dynamical variables to the system coordinates and momenta, to control temperature and pressure. An atom group scaling integrator algorithm was used throughout this study. It consists in a five time step r-RESPA (reversible Reference System Propagation Algorithm) integrator to integrate the equation of motion of our systems, with a 12 fs time step [62]. It was combined with smooth particle mesh Ewald (SPME) to handle electrostatic interac-tions and constraints on covalent bonds entailing hydrogen [63]. The SPME parameters were chosen to maintain a relative error on the electrostatic interaction below 0.1 %. For this purpose, a converge parameter = 0.43 Å$^{-1}$ was used. For all systems a 5$^{th}$ order B-spline took care of the SPME charge interpolation. A 96-point grid in each Cartesian direction was employed for all micelles except for the protein in bulk water where a 64-point grid was adopted instead. As in [17] all our systems were simulated in a periodically replicated primitive body center cubic (bcc) box corresponding to a system with truncated octahedral boundary conditions. Finally, for each system (i.e. protein and micelles), we saved the atomic coordinates once every 240 fs for subsequent analyses.

## III. Results

**A. Structural Properties of the Reverse Micelles.** We first focus on the shape and size of the two aggregates containing the confined protein. In Figure 3, we present the final snapshots (i.e. $t_{sim}$ =10 ns) for the CW-M1 and CW-M2 systems. It can be seen the two reverse micelles are not spherical but present an ellipsoidal shape. To examine this behavior, it is useful in MD to compute the average ratio a/c between the major (*a*) and the minor (*c*) semi-axes lengths of the both micelles and their water core from the inertia tensor of an ellipsoid with an identical mass [9]. For this purpose, we have computed at each point of the MD trajectories, the relative magnitudes of the three principal moments of inertia of the micelles, $I_1 > I_2 > I_3$ and deduced the three semi-axe lengths *a*, *b*, *c* for an aggregate with an ellipsoid geometry with the following relations:

$$I_1 = \frac{1}{5}M_T(a^2 + b^2)$$
$$I_2 = \frac{1}{5}M_T(a^2 + c^2)$$
$$I_3 = \frac{1}{5}M_T(b^2 + c^2)$$

(1)

Where $M_T$ is the total mass of the micelle or the water core (i.e. including only the water molecules). As shown in Figure 4, the a/c values of the two micelles remain stable after typically 3 (CW-M1) and 4 nanoseconds (CW-M2) of production. For the both micelles, the water pool is found more ellipsoidal than the micelle as previously observed for peptide-devoid [9,16] or peptide-containing micelle [17]. The average values of *a/c* computed after discarding the first 3 and 4 ns of the runs, reported in Table 2, are close to 1.25 ± 0.02 and 1.34 ± 0.03 (CW-M1) and 1.41 ± 0.04 and 1.50 ± 0.06 (CW-M2) for the micelle and the water core, respectively. It is difficult to compare our results with experimental or MD studies since to the best of our knowledge, we did not find in literature reports concerning the shape changes of small AOT RM, after incorporation of cytochrome c. Nevertheless, in the past few years, several experimental studies have been carried out to examine the structure of AOT RM after different size proteins and charge incorporation [6-8]. Recently, Gochman-Hecht and Blanco-Peled [64] and Kim and Dungan [65] have examined by SAXS measurements, the structural changes (in size and shape) of confined globular proteins: lysozyme (pI = 11.1) and bovine serum albumin (pI=4.3). They showed that the micellar shape transition (sphere to cylinder micelle) depends on protein size, charge and RM water content. For example in the case of lysozyme, incorporation of the protein into AOT cylindrical RM induced a shape transition to spherical micelles. It was also shown by [65] that protein incorporation (α-lactalbumin) in AOT RM can change the spontaneous curvature of the surfactant wall and the micellar shape to an ellipsoidal aggregate. Since the above results have been carried out with large micelles ($W_o > 20$), it does not seem appropriate to extrapolate these results to small ones. More recently, Chaitanya et al. [18] have examined by computer simulations, the structure of PFPE self assembled RM with a confined "mini-protein" the 20 residue Trp-Cage, solvated in supercritical carbon dioxide ($scCO_2$). These authors have obtained stable spherical micelles. This result contrasts with our previous results [17] and those presented in this paper, but the discrepancy could be explained by the different type of the surfactant, the nature of solvent or the peptide used by the authors.

To compare the sizes of the two micelles with those obtained by SAXS experiments of [29], we have computed the radius of gyration of the aqueous core, $R_g^w$, by including in the calculation only the $SO_3^-$, hydrogen and oxygen atoms of the AOT headgroup and water. We find $R_g^w$ = 20.1 ± 0.5 Å and $R_g^w$ = 24.2 ± 0.3 Å, for CW-M1 and CW-M2, respectively. These values are close to values of $R_g^w$ obtained for $W_o$ values of 5.5 (18.7 ± 0.1 Å) and 9.1 (23.8 ± 0.1 Å) from the linear fit of $R_g^w$ versus $W_o$ (Figure 2). Moreover, for CW-M2 we remark that the $R_g$ value of the aqueous core increases by 2 Å during the last ~3 ns, in contrast with the behavior of CW-M1 (not shown). This is in relation with the substantial structural changes of confined CYTC within the larger micelle (see below). It is interesting to note that in all cases, the two proteins are well confined within the two RM water cores. Indeed the surface contact by Voronoï construction (see Ref. [66] for description of the algorithm) between the confined protein and the solvent atoms remains close to zero during the simulation times. As suggested by Figures 3 and 5, we also find that the protein in the CW-M1 system is more confined in the water pool than in CW-M2. In addition, we have explored the binding of the surfactant to the protein surface. For this purpose, AOT molecules were selected using a cutoff radius $R_{cut}$ as described in [67]. Briefly, an AOT molecule is considered near a protein residue surface if the distance between the AOT and protein residue atoms is less than $R_{cut}=f(R_{AOT}+R_p)$, where $R_{AOT}$ and $R_p$ are the force field van der Waals radii of the AOT atoms and those of the protein respectively, and $f$ is a coefficient taken arbitrarily equal to 1.1. We find no specific sites for AOT binding in CW-M2, whereas in CW-M1 during the whole length of the trajectory, 4 AOT molecules remain attached to the protein, bound to 8 lysines (K7, K8, K22, K25, K27, K86, K87 and K88. Figure 6-a-b). As expected, in both RM, only negatively charged residues are never covered by AOT. For CW-M2 we also observe that the heme binding site is free of AOT molecules during the entire run length. Finally, we point out that on average 39 % and 14 % of the protein accessible surface is occupied by AOT molecules in CW-M1 and CW-M2, respectively.

**B. Structure and Stability of Proteins in AOT Reverse Micelle.** In this section we turn our attention to the structure of the confined CYTC in the two systems and compare it with the same protein in water and in the crystal. At this end, we examine the root mean square deviation of our trajectories from the X-ray structure of the $C_\alpha$ carbon atoms of the protein, or $X_{rms}$. Figure 7 displays the instantaneous $X_{rms}$ of the three systems as a function of time. It shows that in the smallest RM (CW-M1), CYTC is stable with a $X_{rms}$ value close to 1.4 Å while in the largest RM (CW-M2), $X_{rms}$ increases due to the slight change of the protein secondary structure compared to the crystal structure of the protein. One can also notice that the relatively high value, between 1.6 - 2.5 Å, of $X_{rms}$, for the protein in water (CW; black curve) is strongly related to the force field used in this study. Indeed, an additional simulation performed with CW, under the same thermodynamic conditions and the same protocol (described in section IIA), with another CHARMM force field [38] represented in Figure 5 by a green curve, where the H18 and M80 atoms are explicitly bonded to the Fe atom, is more stable than CW (and even CW-M1). This finding is in agreement the fact that these two bonds play a crucial role in the protein secondary structure stability as suggested by [68-70] and also with the fact that this force field better reproduces the heme moiety than the CHARMM default force field. We have also computed the structural alignment of the protein in CW-M1 and CW-M2 with the PyMOL alignment utility [71]. The RMSD value obtained (2.7 Å) confirms that the protein secondary structure is slightly different in the two RM.

As previously mentioned, circular discroïsm spectra indicate that the α-helix content of CYTC in AOT RM, decreases with the increase of the water content [32,33]. To investigate this aspect, we have examined the protein secondary structure evolution as a function of time with the STRIDE program [23] implemented in VMD [72]. Each time point was obtained by extracting every 10 ps the atomic coordinates of the protein (for a total of 10000 atomic configurations) and by computing the secondary structure elements (not shown). It appears that CYTC secondary structures do not differ significantly from the crystal structure. Indeed, we find that in the three trajectories the secondary structure elements [23] of CYTC are conserved to a large extent, except for one $\alpha_1$helix, (residues 3 - 14) in the larger micelle. This finding implies that structural changes occur mainly in the loop regions of the protein. We also remark that the length of our simulations (10 ns) is probably too short to observe any significant denaturation process of the protein. Recent NMR experiments [73] suggest that in RM at $W_o = 10$, CYTC is substantially unfolded. However, within the time scale of our trajectories, we only observe the first steps of the denaturation process, which may take microseconds to be completed. Partial denaturation of CYTC in CW-M2 may increase the size of the confined protein. We have computed the radius of gyration, of the CYTC ($R_g^p$) as a function of time (not shown) and compared it with CYTC in water. We find that the value $R_g^p$ for CYTC in CW-M2 increases slightly during the last 6 ns of the run, up to a value near 13.7 Å, ~ 4.6

%, higher than the value found for CYTC in CW-M1, or in water (13.1 ± 0.1 Å). Nevertheless, these values remain close to those found for CYTC in water by [74] using SAXS (13.8 ± 1.0 Å).

Protein flexibility was also examined as function of the residue number by computing the root mean square fluctuations (RMSF) from the last 4 ns of the simulations (Figure 8). For CW, a major peak is present in loop $L_1$ (residues 15 - 38), showing a larger degree of fluctuation as compared to the rest of the protein. This peak is also present in CW-M2 with two additional peaks in two loops: the $L_2$ loop (residues 40 - 50) and the $L_4$ loop (residues 74 - 88). These three regions contain residues (H18, G41, Y48, T49, W59, T78, K79 and M80) which are hydrogen bonded to the propionate oxygen atoms, bound to the heme iron, and participate to the stabilization of the native heme state [34]. For CW-M1, the RMSF value for each residue is smaller than ~ 0.4 Å$^2$ indicating that the protein is in a "frozen-like" state in agreement with the $X_{rms}$ curve obtained for CW-M1. These results suggest therefore that the heme environment is not conserved when the protein is confined in the large micelle, which is in good agreement with previous spectroscopic findings [29,30].

**C. Structure of Heme.** To gain insights on the structure of heme, we have compared the average distances between the hydrogen bond donors and acceptors of several residues, the oxygen propionate and the Fe atoms of CYTC in AOT RM, in the crystal form of CYTC and in water. These values are obtained from the last 4 ns of the three MD are reported in Table 4 and illustrated in Figures 8(a-d). We are aware that an hydrogen bond is generally characterized by a geometrical criterion such as the angle and distance between the hydrogen bond donor and acceptor, (see for example [75]), but as in ref. [34], we have have used only a distance criterion between the two partners to determine the existence of the hydrogen bond. We first notice that the differences in the heme hydrogen bond distances are small between the X-ray structure and the CW simulation. Comparison between a, b, c, d in Figure 9, clearly shows a partial distortion of the heme pocket in the largest micelle. For CW-M2 we find a large increase in the length of the hydrogen bonds involving residues G41, Y48, W59, K79 and T78, whereas in CW-M1 these bonds are unaffected with respect to CW. Differences between CYTC in RM and the protein in solution are more important for hydrogen bonds involving T49 and N52. It is interesting to note that the breaking of the hydrogen bond between residue W59 and the propionate oxygen atoms of the heme, with the increase of the water content in the AOT RM, has been also observed by alteration of circular dichroïsm and florescence spectra [29]. Indeed circular dichroism spectra have shown that the 282-288-nm bands due to the tryptophan residue orientation and environment disappear with the incorporation of the protein into RM, at a large water content. Moreover, in fluorescence experiments, the low fluorescence intensity at 330 nm at $W_o < 10$ is attributed to the quenching of the excited tryptophan (W59) by the heme [76]. With the increase of $W_o$, the fluorescence intensity of W59 increases and suggests therefore that at $W_o > 10$ the tryptophan ring is now too far from heme to be quenched [77]. Our results are consistent with these results.

The direct heme bond M80-Fe is also found much longer in CW-M2, i.e. 10.9 Å, compared to CW-M1, 3.9 Å, and CW, 3.0 Å. This result is in agreement with spectroscopic data [29,30], which indicate the absence of the 695 nm absorption band (paralleled with the Met80-Fe bond disruption) observed at any water content ($W_o < 10$). Similarly, in both micelles the H18-Fe bond is around 1.4 Å longer than in solution where its length is 3.0Å. Altogether these results show that the heme pocket is more deformed and open in the largest micelle than in the smallest or in solution.

**D. Micellar Water Dynamics.** Experimental techniques [57,78-81] and simulations [12,13,14,16] have shown that in AOT RM, the micellar water properties differ substantially from those of bulk water. In particular it was noted that the translational and rotational properties of the micellar water are slowed down in small AOT RM ($W_o < 10 - 15$) as compared to those in the bulk. Indeed as the water amount in the RM water pool increases, the dynamics of confined and bulk water get more similar. To examine this aspect, we have computed the mean square displacement (MSD), $<|r(t)-r(0)|^2>$, of the core water in protein-void and protein-filled RM as a function of time, and compared it with the results obtained with TIP3 bulk water. These functions are displayed in Figure 10. As previously observed for water molecules near protein surfaces [82,83], micelles [84] and in AOT reverse micelles [9,13], their MSD is smaller than that of bulk water and presents a subdiffusive regime which obeys a power law (i.e. $\langle |r(t)-r(0)|^2 \rangle \propto t^\alpha$ with $\alpha < 1$ denoting a dispersive diffusion. We observe the slowest diffusion for the smallest micelle (CW-M1), whereas in the largest one the water diffusion is similar to that of protein-void micelles. From these curves, we estimate the water residence time, $\tau_w$, 32.6 ps and 12.0 ps for CW-M1 and CW-M2, respectively. In contrast, we find values of 7.5 ps and 6.5 ps for protein-void micelles at $W_o = 5$ and 7 [9], respectively. Following Denisov and Halle [85], we define $\tau_w$ as the time needed by a water molecule to cover a distance of 3 Å, i.e. corresponding to a water molecular diameter. Our calculations show that water in the first hydration shell around the protein is slowed down to a large extent with respect to water hydrating the protein surface in the bulk. Compared with $\tau_w = 13$ ps for CYTC, we find a $\tau_w$ of 113 ps and 59 ps for CW-M1 and CW-M2, respectively. In a previous work with an alanine octapeptide ($A_8$) confined in AOT RM [17], we showed that the secondary structure of the peptide is stable in a smaller RM, when all water present is assumed to be bound to the AOT head groups (and $Na^+$ ions). We interpreted this behavior by observing that the stability of the $A_8$ peptide in the small AOT RM is enhanced by the effect of hydration competition between the surfactant head groups and the peptide polar groups, as well as by the confinement effect due to the small size of the water pool, favoring proximity between the surfactant headgroups and the protein surface (Figures 3 and 5). We notice that the slow dynamics observed in the water confined within the smallest micelle CW-M1, may also contribute to a significant extent to the protein stability. Furthermore, this effect illustrates the competition for water existing between the protein and the micellar wall and can thus be considered as a marker of the observed changes.

## IV. Conclusions

In this paper, we have examined by computer simulation the confinement of the reduced form of CYTC in AOT RM, at two levels of hydration. In AOT reverse micelles, strong electrostatic interactions occur between the anionic polar head groups and the positively charged protein residues (pI=10.5). The spectroscopic results previously reported by [29] have been interpreted as originating from the opening of the CYTC heme crevice under the electrostatic field generated by the micellar anionic head groups, since most of the CYTC positive charges are located around the heme crevice [25-27]. Our simulations support this picture. In CW-M1, where the size of the micelle is close to that of the protein molecule, our simulations display within its time window (10 ns), a confined, rather stable protein, despite the binding of four AOT molecules on the lysine residues. At a higher hydration level, we do observe a heme pocket distortion starting at $W_o = 9.1$, a lower value compared to the value of $W_o = 20$, reported by [29]. This distortion involves a partial opening of the prosthetic group crevice, which becomes accessible to water during the whole run. At the same hydration level we observe large fluctuations of $L_1$, $L_2$, and $L_4$, the loops involved in the stabilization of heme native state. The CYTC peptide chain itself exhibits minor changes, with the exception of one α-helix (residues 2-15). It is clear that several parameters account for the observed structural changes, the micelle-protein interactions constituting the major factor, compared to the membrane-protein interactions which disrupt the native protein structure, [26,28] including the disruption of the Met-80 ligation to the iron atom.

In contrast, the observed slowing down of water molecules at the protein surface can be considered a marker of the microenvironment experienced by water molecules. In this respect, we stress that the dynamic slowing down increases at low water content, where the protein distortion is minimal. Thus, our simulations provide clues that the structural alterations observed at heme level are the first signals of CYTC denaturation, observed experimentally at high water content. Such unfolding may take microseconds to be completed in AOT RMs, a time too long to be explored by classical explicit molecular simulations.

**Acknowledgements:** We thank the Centre de Calcul Recherche et Technologie (CCRT) in Bruyères le Châtel for the generous allocation of computation time used for this project.

**Tables**

| System | CW | CW-M1 | CW-M2 |
|---|---|---|---|
| $W_o$ | - | 5.5 | 9.1 |
| $n_{AOT}$ | - | 106 | 147 |
| $n_{Na+}$ | - | 106 | 147 |
| $n_{Cl^-}$ | 6 | 6 | 6 |
| $n_{H2O}$ | 6681 | 580 | 1339 |
| $n_{iso}$ | - | 2587 | 3345 |
| $n_{atoms}$ | 21796 | 77751 | 102338 |
| $m_{AOT}/m_{iso}$ (%) | - | 83.7 | 81.1 |
| $t_{sim}$(ns) | 10 | 10 | 10 |

**Table 1.** System composition. $n_x$ is the number of x molecules in the system; $t_{sim}$, the simulation time excluding the equilibration periods 100 ps for CW and 200 ps for CW-M1 and CW-M2.

| System | | $<a/c>$ | $R^w_g$ |
|---|---|---|---|
| CW-M1 | Micelle | 1.25 ± 0.02 | 22.3 ± 0.1 |
| | Aqueous Core | 1.34 ± 0.03 | 20.1 ± 0.5 (18.7 ± 0.1) |
| CW-M2 | Micelle | 1.41 ± 0.04 | 26.5 ± 0.1 |
| | Aqueous Core | 1.50 ± 0.06 | 24.2 ± 0.3 (23.8 ± 0.1) |

**Table 2.** Average reverse micelle and aqueous core shapes and dimensions, computed from the last 7 and 6 ns for CW-M1 and CW-M2 of the runs, respectively. The quantities labeled Micelle and Aqueous Core were computed by including all atoms of the micelle and those of water and AOT headgroup, respectively. Experimental values (in parenthesis) are calculated from the linear fit of $R^w_g$ as a function $W_o$ (Figure 1).

| System | Crystal | CW | CW-M1 | CW-M2 |
|---|---|---|---|---|
| | | Average Distances | | |
| R38NH1-O1A | 3.42 | 3.60 | 2.95 | 2.87 |
| G41N-O2A | 2.94 | 3.22 | 3.09 | 5.81 |
| Y48OH-O1A | 2.36 | 2.76 | 2.66 | 8.01 |
| T49OG1-O1D | 3.12 | 3.69 | 6.22 | 7.47 |
| T49N-O2D | 2.93 | 3.62 | 5.39 | 7.22 |
| N52ND2-O2A | 3.22 | 3.32 | 4.46 | 4.17 |
| W59NE2-O2A | 2.74 | 2.83 | 3.01 | 4.65 |
| T78OG1-O1D | 2.56 | 3.61 | 2.58 | 6.25 |
| K79N-O1D | 3.06 | 3.34 | 4.00 | 6.75 |
| M80SD-FE | 2.30 | 3.04 | 3.90 | 10.89 |
| H18NE2-FE | 2.16 | 3.00 | 4.20 | 4.45 |

**Table 4.** Distances in Å for selected residues atoms from the oxygen propionate and the iron of the heme. See Figures 8 and 9 for the localization of each atom and residue in the protein backbone.

| System | CW-M1 | CW-M2 | Bulk |
|---|---|---|---|
| $\tau_w$ | 32.6 | 12.0 | 2.4 |
| $\tau_w/\tau^b_w$ | 13.6 | 5.0 | 1.0 |

**Table 5.** Translational diffusion of the total water confined in AOT reverse micelles and in the bulk. $\tau_w$ is the water residence time, defined as the time (in ps) for a water to cover its own diameter (i.e. 3 Å) [85]. $\tau_w/\tau^b_w$ is the ratio between the water residence time in the micelle and in bulk water and expresses the retardation compared to bulk water (see main text).

**Figures**

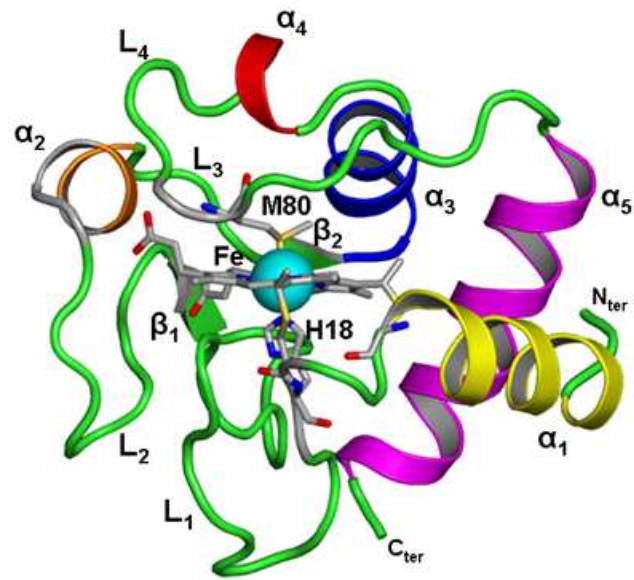

**Figure 1.** Crystal structure of horse cytochrome C (CYTC) (PDB code: 1HRC) exhibits 45 % helical ($\alpha_{1-5}$), 5 %, 2 β sheets ($\beta_{1-2}$) and 4 majors loops ($L_{1-4}$). H18 and M80: Histidine 18 and Methionine 80. Secondary structure was delimited with the STRIDE program [23] and according to [34].

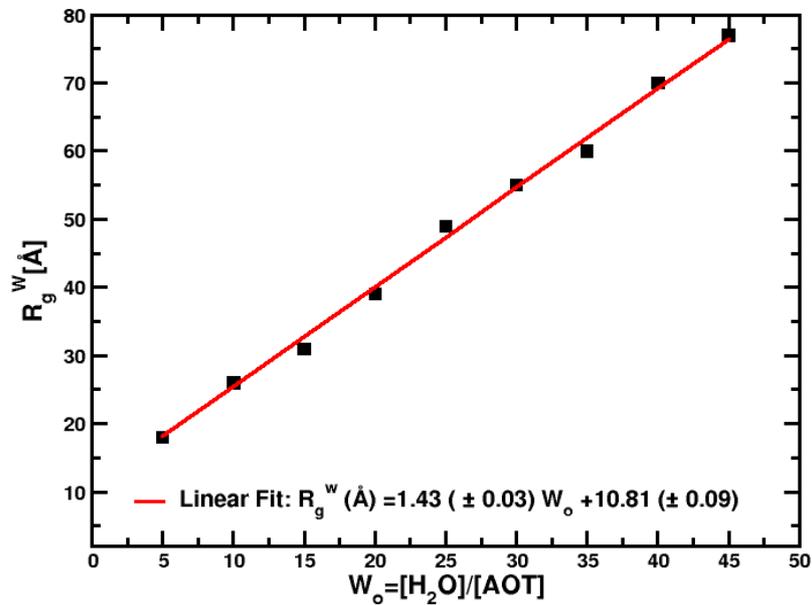

**Figure 2.** Plot of the radius of gyration of the aqueous core (AOT headgroup and water) of the reverse micelle, $R_g^w$ as function of $W_o$ obtained from SAXS [29] with [AOT] = 0.1 M and [CYTC] = 0.01 M at 284 K.

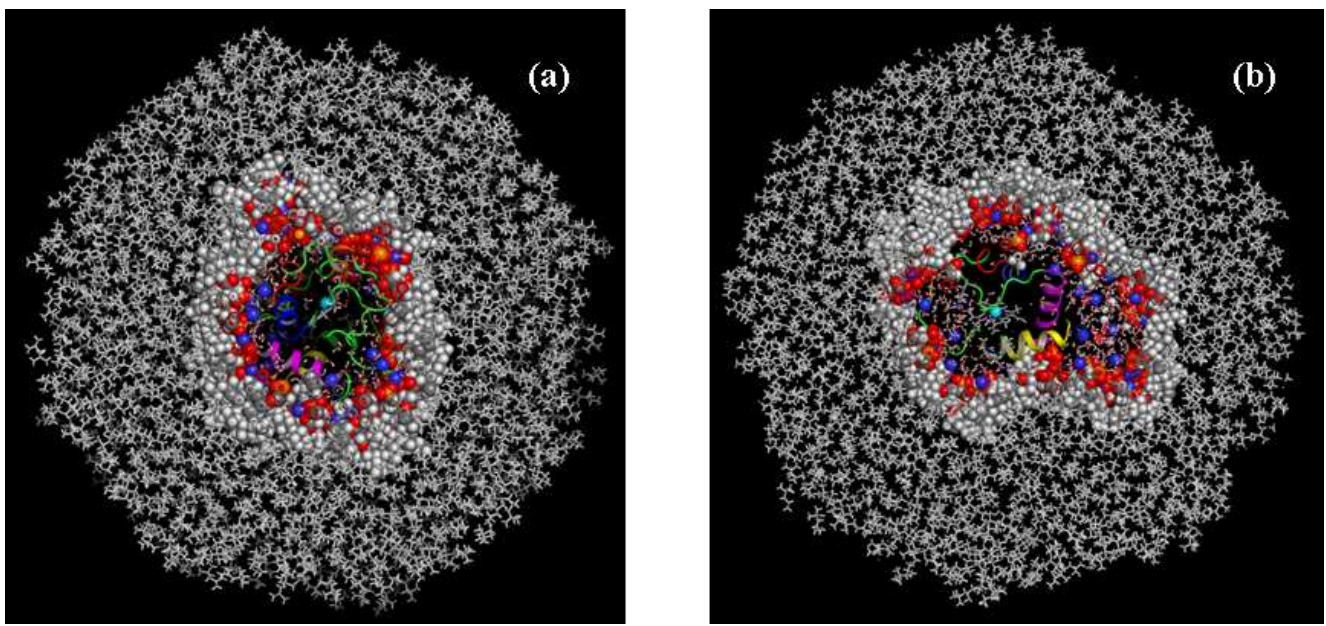

**Figure 3.** Final configurations ($t_{sim}$ = 10 ns) of CW-M1 (a) and CW-M2 (b). The atoms of AOT, water, sodium and isooctane molecules are represented by spheres and sticks; the confined protein in cartoon representation. Figures produced with PyMol [71].

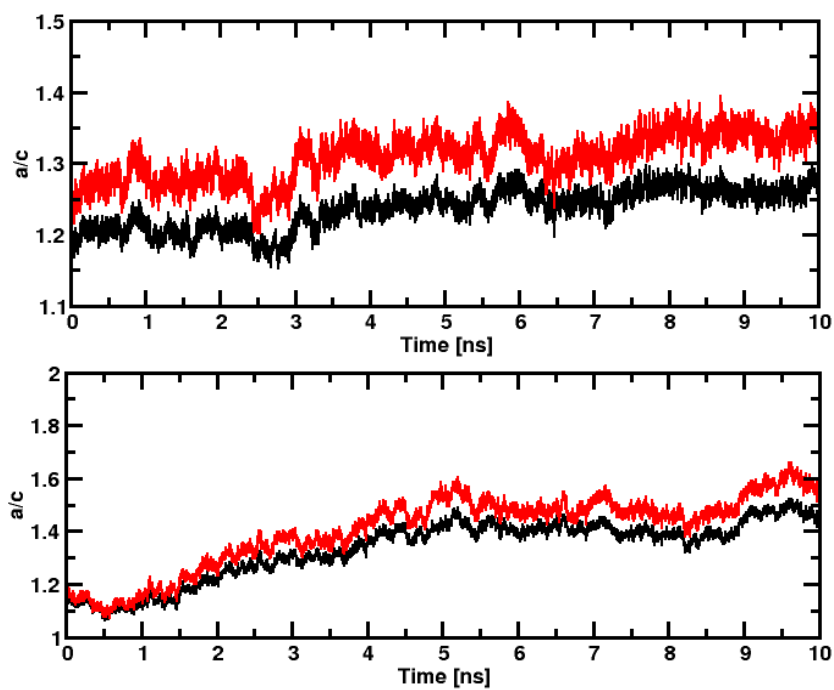

**Figure 4.** Semi-axial ratio a/c, of the whole micelle (black) and water core (red) as function of time. CW-M1 (top) and CW-M2 (bottom).

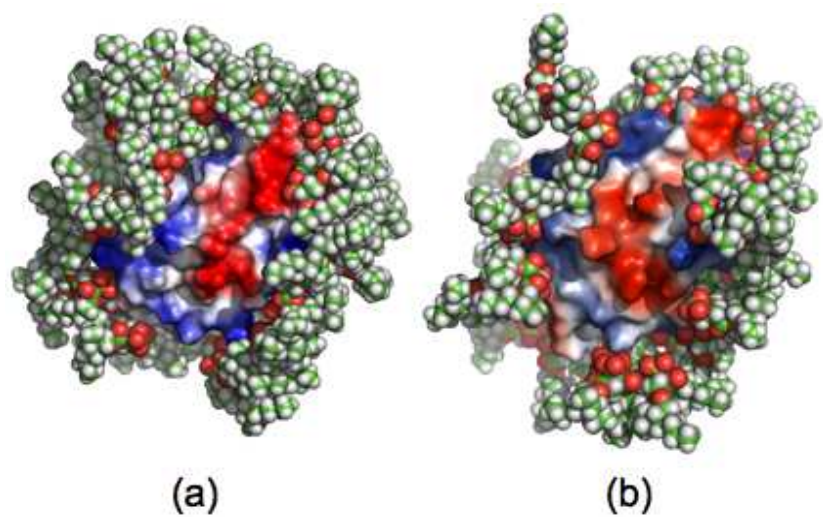

**Figure 5.** Snapshots of the micelles CW-M1 (a) and CW-M2 (b) at the end of the run. Hydrophobic and hydrophilic protein residues accessible to AOT surfactant headgroup are plotted in red and blue, respectively.

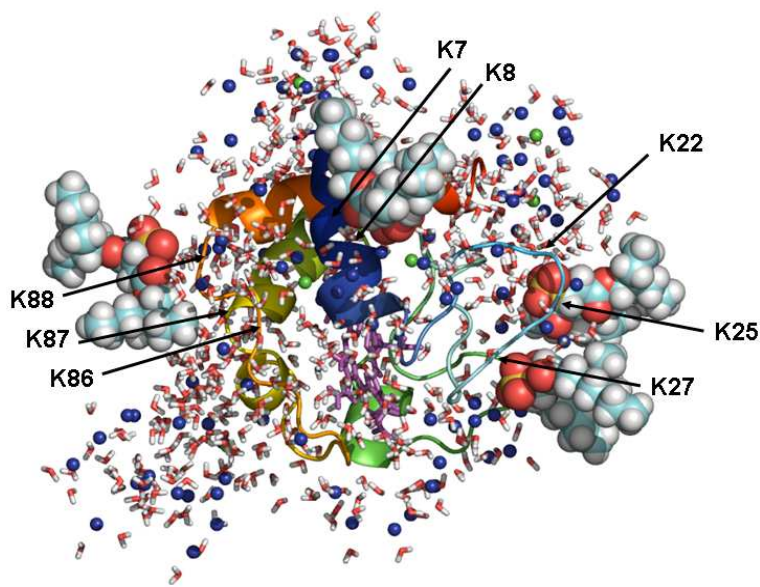

**Figure 6**. Localization of the 4 permanently bound AOT molecules around the 8 lysines of the protein in CW-M1. The AOT molecules were selected using a cutoff radius $R_{cut}$ (see main text for details).

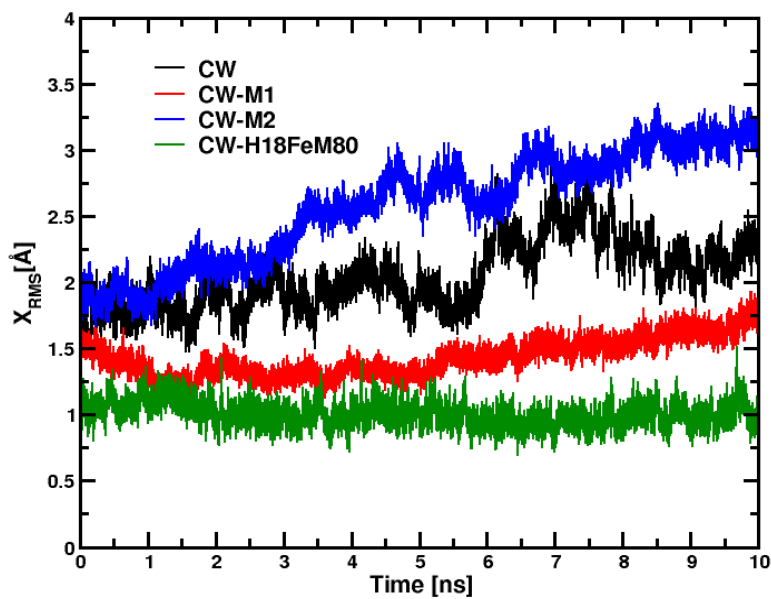

**Figure 7.** Computed instantaneous root mean square deviation from CYTC X-ray structure, or $X_{rms}$, of the three systems as a function of time. The green curve is obtained from an additional MD with the M80-Fe and H18-Fe explicitly bonded (see main text for details).

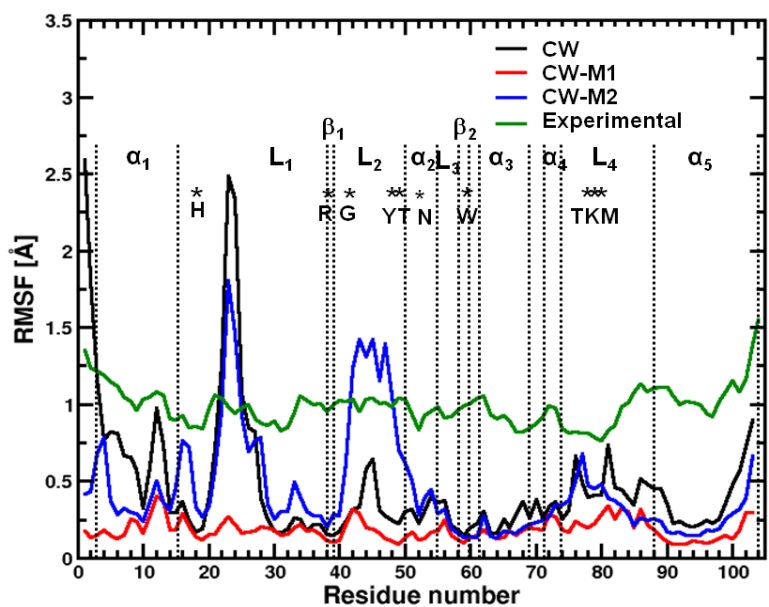

**Figure 8.** Root Mean Square Fluctuations (RMSF) computed for each residue of the protein for the last 4 ns of the simulations. Experimental values, in green, are taken from the X-ray B-factor. Only $C_\alpha$ backbone atoms are considered. Residues bonded to heme are shown with a star and one code letter. The dashed lines represent limits of calculated secondary structures using STRIDE, according to Figure 1. Computing the RMSF with different procedures (i.e. averaged structures and fluctuations computed in different regions of the trajectories) gives differences of at most 1.5 Å$^2$ for the highest peaks, but much smaller, within the order of ~ 0.1 Å$^2$, in the other regions of the plot.

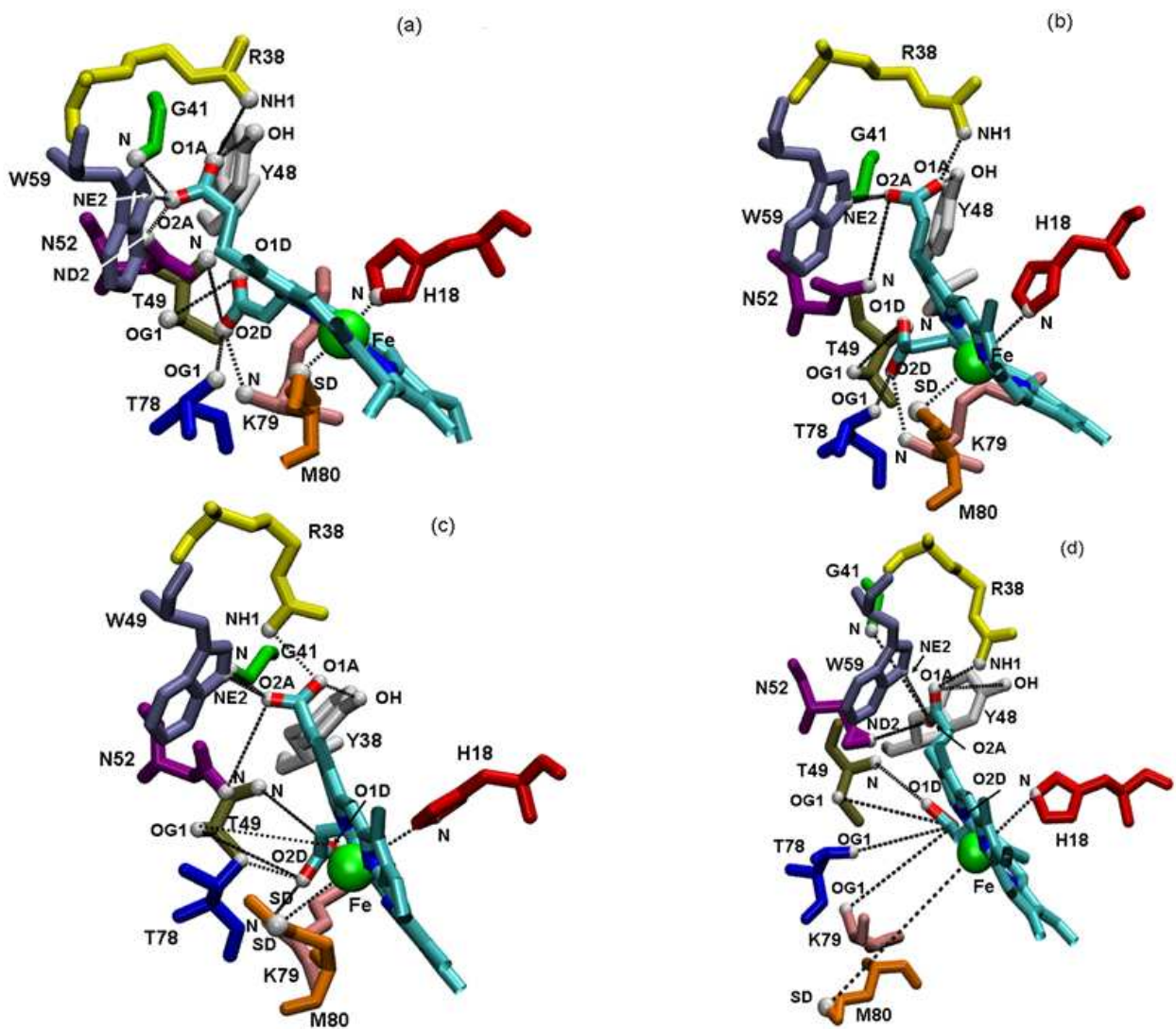

**Figure 9.** Average conformations of the heme environment for the last 4 ns of the simulations of each protein: Crystal (a), CW (b), CW-M1(c) and CW-M2 (d). Hydrogen atoms are not shown for visual clarity and atom types are labeled according to the CHARMM force field. Distances between the hydrogen bond donor and the propionate atoms of the heme and the iron are drawn as dashed lines. These pictures were produced with pyMOL [71].

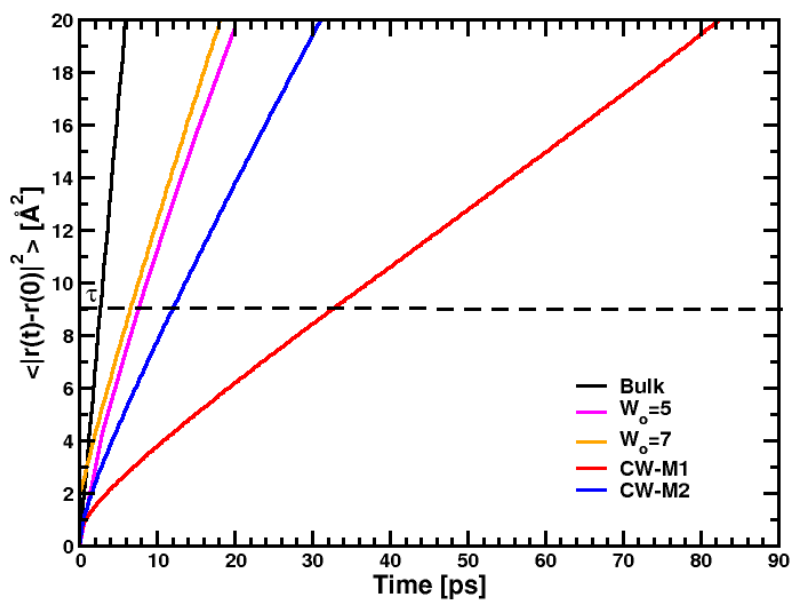

**Figure 10.** Water mean square displacement (msd) as function of time. The dashed line indicates a msd of 9 Å. The RSD for two "empty" micelles, $W_o$=5 and $W_o$=7 are taken from Ref. [9].

**COVER PICTURE FOR THE PAPER**

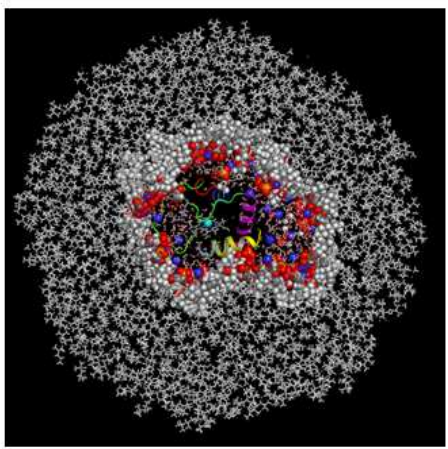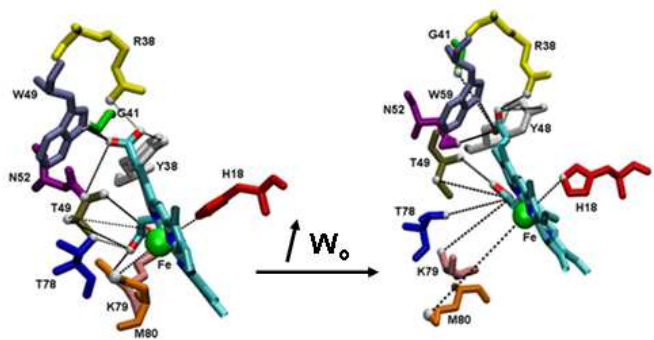